\documentclass[superscriptaddress,twocolumn,aps,prb,floatfix,showpacs,amsmath,amssymb,superbib]{revtex4}
\usepackage{epsfig}
\usepackage{dcolumn}
\usepackage{bm}
\begin{document}
\title{Non-Gaussian Noise in the In-Plane Transport of Lightly Doped La$_{2-x}$Sr$_x$CuO$_4$: Evidence for a Collective State of Charge Clusters}

\author{I. Rai\v{c}evi\'{c}}
  \affiliation{National High Magnetic Field Laboratory and Department of Physics, Florida State University, Tallahassee, FL 32310, USA}
\author{Dragana Popovi\'c}
\email{dragana@magnet.fsu.edu}
\affiliation{National High Magnetic Field Laboratory and
Department of Physics, Florida State University, Tallahassee, FL 32310, USA}
\author{C. Panagopoulos}
\affiliation{Department of Physics, University of Crete and FORTH,
71003 Heraklion, Greece} \affiliation{Division of Physics and
Applied Physics, Nanyang Technological University, Singapore}
\author{T. Sasagawa}
\affiliation{Materials and Structures Laboratory, Tokyo Institute
of Technology, Kanagawa 226-8503, Japan }
\date{\today}

\begin{abstract}
A study of the in-plane resistance noise on a high quality single crystal of La$_{1.97}$Sr$_{0.03}$CuO$_{4}$ deep inside the spin-glass phase reveals the onset of non-Gaussianity and the insensitivity of the noise to the in-plane magnetic field.  The results indicate that the charge dynamics becomes increasingly slow and correlated as temperature $T\rightarrow 0$.  The analysis of the higher order noise statistics provides evidence for the existence of a collective ground state of charge clusters, which seem to coexist with charge-poor antiferromagnetic domains that are frozen at such low $T$.

\end{abstract}


\maketitle
\section{INTRODUCTION}

There is growing experimental evidence for the emergence of spatial inhomogeneities at the nanoscale in many strongly correlated electron systems.  In addition to charge, other degrees of freedom, such as spin and lattice, typically also play an important role in these materials, leading to the competition of several ground states and the resulting nanoscale phase separation.\cite{Gorkov,Kivelson93,Elbio}  Such nanophases have been variously described as bubbles, stripes, clumps, clusters, or domains, depending on system details.  Other investigations have discussed the emergence of inhomogeneous charge-ordered phases in analogy with the smectic and nematic states in liquid crystals.\cite{Kivelson98}  In general, many different configurations of such nanoscopic ordered regions often have comparable (free) energies, such that these metastable states are separated by barriers with a wide distribution of heights and, thus, relaxation times.  This leads to the slow dynamics typical of glassy or out-of-equilibrium systems.  In fact, the frustration caused by the competition of interactions on different length scales may give rise to the emergence of an exponentially large number of metastable configurations and the associated glassy dynamics even in the absence of disorder.\cite{Schmalian00}  This scenario is relevant to many materials, including cuprates, where a long-range Coulomb repulsion competes with a short-range attraction that results from magnetic exchange interactions.  The disorder should further stabilize the glassy phase.  Even though the emergence of glassiness thus appears to be almost ubiquitous at low temperatures ($T$), the glassy charge dynamics and out-of-equilibrium behavior in general remain poorly understood.

In cuprates, various experimental techniques provide evidence for glassiness in the spin sector at $T\leq T_{SG}(x)$, where $T_{SG}$ is the spin glass transition temperature and $x$ is the doping.\cite{Chou95,Niedermayer98,Julien99,Wakimoto00,Christos-SG}  In La$_{2}$Sr$_{2-x}$CuO$_{4}$ (LSCO), where the long-range antiferromagnetic (AF) order of the parent compound La$_2$CuO$_4$ is completely suppressed for $x\approx 0.02$ but two-dimensional short-range AF correlations persist\cite{Kastner98}, the high-temperature superconductivity (HTSC) sets in at $x\approx 0.055$.  The spin glass (SG) phase emerges with the first added holes and extends all the way to slightly overdoped $x\simeq 0.19$ (Refs.~\onlinecite{Niedermayer98,Christos-SG}), thus coexisting with HTSC.
Unlike conventional spin glasses, here the low-$T$ phase results from cooperative freezing of moments in different domains in which spins are antiferromagnetically ordered.\cite{Cho92}  Therefore, the spin glass has a local ``stripe'' character and is often called a ``cluster spin glass''.  On the other hand, other experiments suggest that charge is clustered in antiphase boundaries \cite{Matsuda1,Matsuda2,Wakimoto00,Lavrov01} that separate the hole-poor AF domains in CuO$_2$ ($ab$) planes.\cite{Julien99,Singer02NQR,Dumm03EM,Ando02Ranisotropy,Ando03MR}  Imaging techniques on other cuprates (see, \textit{e.g.}, Ref.~\onlinecite{kohsaka07}) have confirmed the presence of randomly distributed charge domains with short-range order in the underdoped or pseudogap regime.  It is plausible to expect, based on general arguments, that the ground state of such a random distribution of nanoscale charge domains is a charge glass.  The nature of the ground state is clearly of great interest, because it is from this state that the HTSC emerges with hole doping.  However, studies of charge glassiness, \textit{i.e.} of  \textit{dynamic} as opposed to static charge ordering, have been relatively scarce.\cite{Thompson-glass}  Moreover, experiments at relatively high $T$ may be difficult to interpret in many materials because of the changes in structural or magnetic symmetry.  At low $T$, on the other hand, the charge dynamics may become so slow that the charge distribution appears to be static on experimental time scales.

Therefore, the resistance noise spectroscopy employed at very low $T$ and on very long time ($t$) scales is clearly a technique that is well suited for probing the charge dynamics.  It has already been used successfully in charge (Coulomb) glasses in nonmagnetic systems.\cite{Bogdanovich02,Jaroszynski02,Jaroszynski04}  The analysis of the noise statistics can provide evidence for the presence of a large number of metastable states, which are an essential prerequisite for well-known dynamical features of glassiness, such as aging, memory, and breaking of ergodicity.

\begin{figure}
\includegraphics[width=6cm]{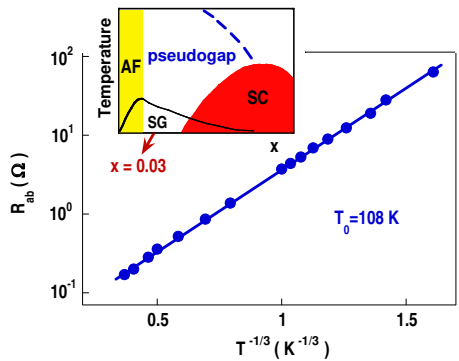}
\caption {(Color online) The temperature dependence of the zero-field cooled in-plane resistance $R_{ab}$ below 30~K (Ref.~\onlinecite{Ivana-pMR}).  The solid line is a fit with slope $T_0$.  Inset: Schematic phase diagram of hole-doped LSCO.
}\label{fig:LSCOabRvsTPhaseDiagram}
\end{figure}

In LSCO, the noise has been measured in the $c$-axis or out-of-plane resistance $R_c$ of the $x=0.03$ sample,\cite{Raicevic08} which is located in the pseudogap regime, but it does not superconduct at any $T$ (Fig.~\ref{fig:LSCOabRvsTPhaseDiagram} inset).  The data have provided evidence that, deep within the SG phase at $T\ll T_{SG}$ where transport is insulating, the charge dynamics become increasingly slow and correlated, \textit{i.e.} glassy, as $T\rightarrow 0$ (Ref.~\onlinecite{Raicevic08}). Moreover, the results strongly suggest that doped holes form a cluster charge glass.  These conclusions are further supported by
%
%
impedance spectroscopy.\cite{Jelbert08}

In the same regime where $R_c$ noise shows evidence for charge glassiness, both $c$-axis\cite{Raicevic08,Ivana-pMR} and in-plane magnetoresistance\cite{Ivana-pMR} (MR) exhibit the emergence of a strong, positive component for all orientations of the magnetic field $B$.  This positive MR (pMR) shows a hysteresis and memory, indicative of the presence of correlated magnetic domains.  A careful and comprehensive analysis of the data\cite{Raicevic08,Ivana-pMR} points to the picture of AF domains that are frozen at low $T\ll T_{SG}$, and holes confined to the domain ``walls'' (\textit{i.e.} patchy areas separating the AF domains).  While low $B$ produces some motion of the domain walls leading to a hysteretic MR, the main transport mechanism that gives rise to the pMR remains unchanged and, in fact, is the same as that observed in various nonmagnetic, disordered materials with strong Coulomb interactions.  Basically, in a system that conducts via variable-range hopping (VRH), the Zeeman splitting in the presence of a Coulomb repulsion between two holes in the same disorder-localized state leads to a pMR by blocking certain hopping channels.\cite{KK,Meir}  Much higher $B$ leads to the reorientation of the weak ferromagnetic (FM) moments\cite{Thio88,Thio90} associated with each AF domain and oriented along the $c$ axis, resulting in a negative MR.\cite{kotov07}  Therefore, since the magnetic background for a fixed $B$ is frozen at $T\ll T_{SG}$,
at least on experimental time scales, only holes in the domain walls contribute to transport and glassiness in the resistance noise.  For this reason, the statistics of the $R_c$ noise, which reflects a collective wandering of the system among many metastable states, is unaffected by the presence of a fixed $B$, and it is also independent of the magnetic history.\cite{Raicevic08}

While the previous study of the $R_c$ noise in La$_{1.97}$Sr$_{0.03}$CuO$_{4}$ strongly suggests\cite{Raicevic08} that the doped holes form a cluster glass state as a result of Coulomb interactions, albeit in the presence of a frozen, random magnetic background, the noise in the in-plane transport has not been investigated.  In addition, only the effect of $B\parallel c$ was explored.  However, the in-plane transport is generally agreed to be more relevant to the physics of cuprates and, in particular, to the emergence of HTSC at higher charge-carrier concentrations.  Furthermore, various theoretical models, such as those that consider fluctuations of the local electronic order and the effect on transport noise,\cite{kivelson06,STM-noise} have focused on the in-plane transport and in-plane symmetry breaking.  It should be noted that the relationship between the in-plane and out-of-plane transport in these highly anisotropic materials is far from trivial, and their sometimes contrasting behavior has been one of the most unusual characteristics of the cuprates.  Indeed, comparative studies of intralayer and interlayer properties continue to provide key insights into the physics of these strongly correlated systems.\cite{stripes-NJP}

Therefore, here we investigate the noise in the in-plane resistance $R_{ab}$ of La$_{1.97}$Sr$_{0.03}$CuO$_{4}$ at very low $T\ll T_{SG}$ and with $B\parallel ab$.  Just like what was observed\cite{Raicevic08} for the $c$-axis transport and a different field orientation ($B\parallel c$), the fluctuations of the in-plane resistance $R_{ab}$ with time provide evidence for the increasingly slow, correlated dynamics as $T\rightarrow 0$.  Here, however, the magnitude of the resistance noise is significantly smaller and the emergence of charge glassiness takes place at even lower $T$ than for the out-of-plane transport.  In spite of these quantitative differences, the overall data strongly support earlier conclusions\cite{Raicevic08,Jelbert08,Ivana-pMR} and provide evidence that doped holes in lightly doped LSCO form a collective, glassy state of charge domains or clusters located in CuO$_2$ planes.

\section{EXPERIMENT}

A high quality single crystal of LSCO with $x=0.03$ was grown by the traveling-solvent floating-zone technique.\cite{SasagawaLSCO}  Detailed measurements were performed on a sample that was cut out along the main crystallographic axes and polished into a bar with dimensions $2.10 \times 0.44 \times 0.42$~mm$^{3}$, suitable for direct $R_{ab}$ measurements.  Electrical contacts were made by evaporating gold on polished crystal surfaces, followed by annealing in air at $700^{\circ}$C. For current contacts, the whole area of two opposing side faces was covered with gold to ensure a uniform current flow through the sample. In turn, the voltage contacts were made narrow ($\sim80~\mu$m) in order to minimize the uncertainty in the absolute values of the resistance.  Gold
leads were attached to the sample using Dupont 6838 silver paste. This was followed by the heat treatment at $450^{\circ}$C in the flow of oxygen for 6 minutes.  The resulting contact resistances were less than 1~$\Omega$ at room $T$.

\begin{figure}
\includegraphics[width=5.5cm]{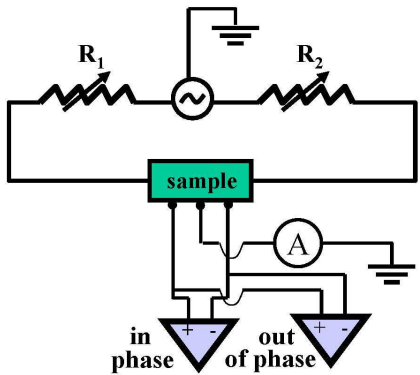}
\caption {(Color online) Schematic of the circuit for the five-point ac bridge
measurement of the voltage (resistance) fluctuations. Ballast resistors $R_{1}$ and $R_{2}$ ($\sim 100$~M$\Omega$) balance the bridge, thus minimizing the effects of temperature and current fluctuations. Two PAR124A lock-in amplifiers detect the in-phase and out-of-phase components of the signal.  The power spectrum of the former contains both sample resistance noise and background, while the power spectrum of the latter contains only the background noise.  The total current was measured with an ITHACO 1211 current preamplifier and a PAR124A lock-in amplifier.
}\label{fig:BRIDGE}
\end{figure}

The sample resistance was measured using the standard four-probe ac technique (typically at $\sim7$~Hz) in the Ohmic regime, at $T$ down to 0.05~K realized in a dilution refrigerator.  The low-frequency noise was measured using a conventional five-probe ac bridge method\cite{Scofield87} with the bridge current $I\approx 100$~nA in the Ohmic regime (Fig.~\ref{fig:BRIDGE}). This method makes the resistance noise measurement nearly insensitive to both $T$ and current fluctuations.  For example, it was confirmed experimentally that the correlation between $T$ and voltage fluctuations decreases and becomes negligible as the balance of the bridge is increased.  Likewise, the correlation between the fluctuations of voltage and current was also negligible.  Two lock-in amplifiers ($\sim 7$ Hz) were used to detect the difference voltage.  This measurement setup makes it possible to remove the contribution of the background noise, which is due to the Johnson noise and noise from the preamplifier.\cite{othernoise}  In this experiment, the power spectrum of the background noise was always negligibly small and white.  (The background noise measured with zero current was white for all $T$ and $B$ and 2--3 orders of magnitude smaller than the sample resistance noise.)  The output filters of the lock-in amplifiers served as an antialiasing device.  Most of the noise spectra were obtained in the $f=10^{-4}-10^{-1}$~Hz bandwidth, where the upper bound was set by the low frequency of $I$, limited by the low cutoff frequency of RC filters used to reduce external electromagnetic noise as well as by the resistance of the sample.

The sample is located in the SG region of the phase diagram (Fig.~\ref{fig:LSCOabRvsTPhaseDiagram} inset) at $T\leq T_{SG}\sim7-8$~K (Ref.~\onlinecite{Panagopoulos05}).  In that $T$ regime, $R_{ab}$ shows insulating behavior.  In particular, it obeys the VRH dependence $R_{ab}=R_{0}\exp(T_{0}/T)^{\mu}$, $T_0=108$~K, with an exponent $\mu =1/3$ all the way up to $T =30$~K (Fig.~\ref{fig:LSCOabRvsTPhaseDiagram}).  ($R_{ab}$ is henceforth denoted by $R$ for simplicity.)  The value $\mu=1/3$ is characteristic of two-dimensional (2D) VRH and it is in agreement with early results on ceramic LSCO samples where it was shown that this exponent is doping dependent.\cite{ellman89}  The localization length obtained from the VRH fit\cite{Ivana-pMR} is $\xi\sim 90$~\AA, which is somewhat larger than the average magnetic correlation length $\sim 40$~\AA\, at this doping.\cite{Kastner98,Wakimoto00}  The most probable hopping distance\cite{shklovskii84} $r_h(T)\sim\xi(T_0/T)^{\mu}$ thus spans on the order of ten AF domains at low $T$.  Therefore, for the VRH process, the system appears to be uniform.

\section{Resistance noise}

\subsection{Noise in zero magnetic field}

\begin{figure}[t]
\includegraphics[width=8cm]{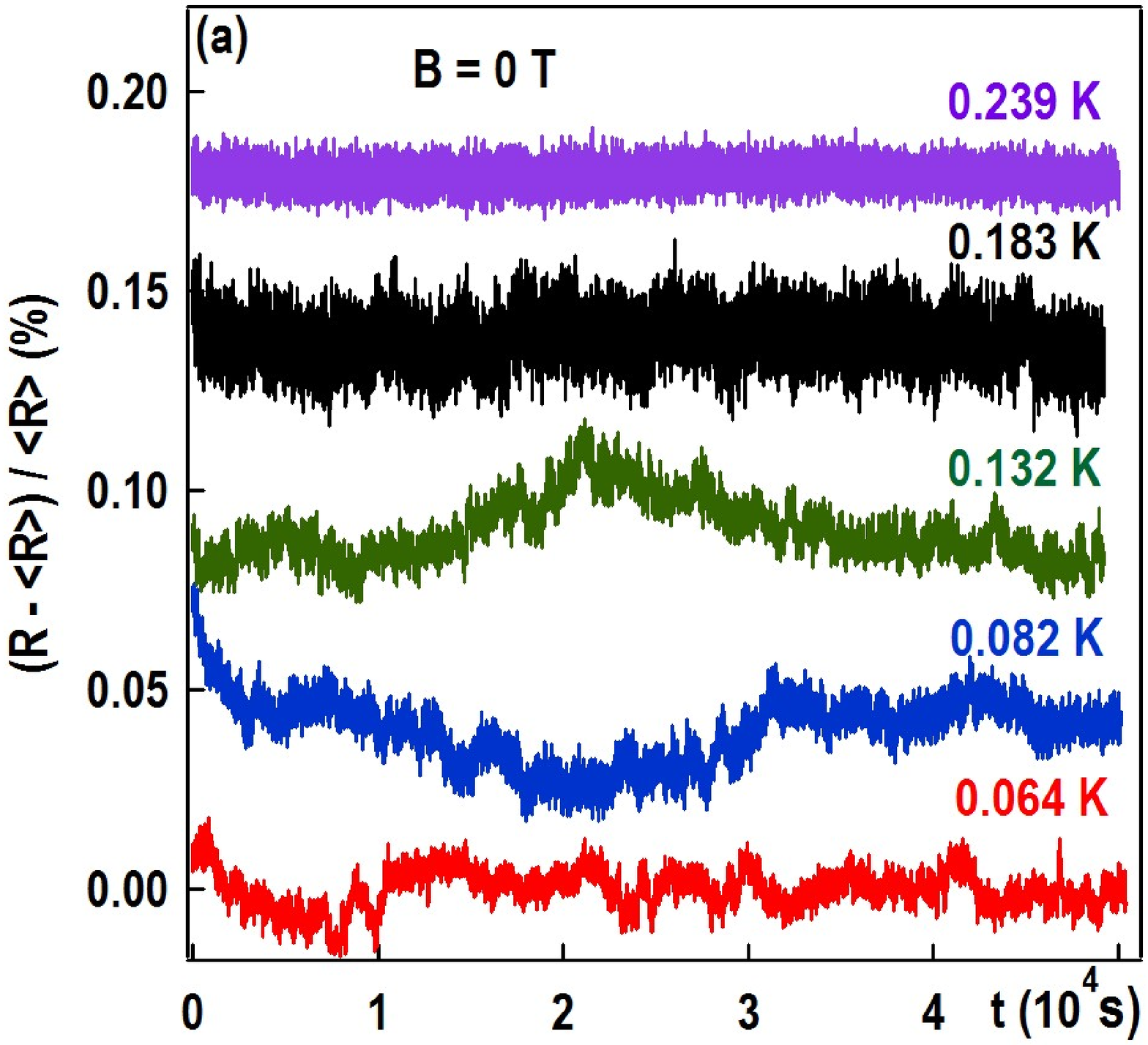}\\
\includegraphics[width=8cm]{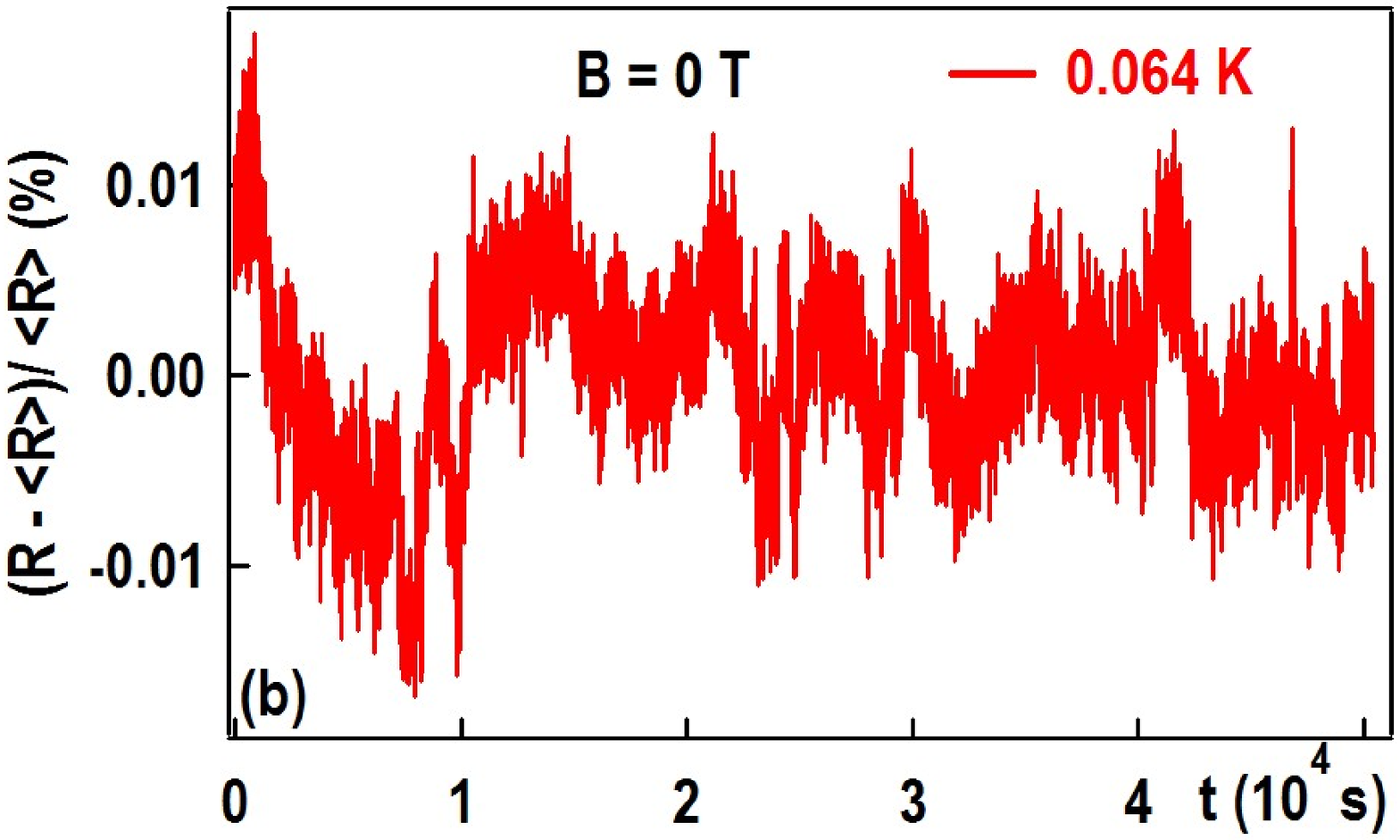}
\caption {(Color online) (a) Relative fluctuations of the in-plane resistance
$\Delta R(t)/\langle R\rangle$ = ($R-\langle R\rangle)/\langle R\rangle$ as a function of time for $B=0$ at several $T$.  $\langle R\rangle$ is the time-averaged resistance determined from a four-probe measurement at each $T$.  All traces are shifted vertically for clarity.  (b) The $T=0.064$~K data in (a) shown on an expanded scale for clarity.}\label{fig:LSCOabNoiseZeroField}
\end{figure}

Figure~\ref{fig:LSCOabNoiseZeroField}(a) shows the
%
%
typical time traces of the relative changes in resistance $\Delta R(t)/\langle R(t)\rangle$ in zero field at various $T$. It is obvious that, at the lowest $T$ [see also Fig.~\ref{fig:LSCOabNoiseZeroField}(b)], the system exhibits random fluctuations on many different time scales, including slow changes over several hours.  While the $R_c$ noise occasionally exhibited also some discrete switching events,\cite{Raicevic08} such fluctuations have not been observed in the in-plane resistance noise measurements.  On the other hand, similar to the $c$-axis transport, here the magnitude of the noise does not appear to depend on $T$, but the character of the noise is significantly altered with increasing $T$.

\begin{figure}[t]
\includegraphics[width=7cm]{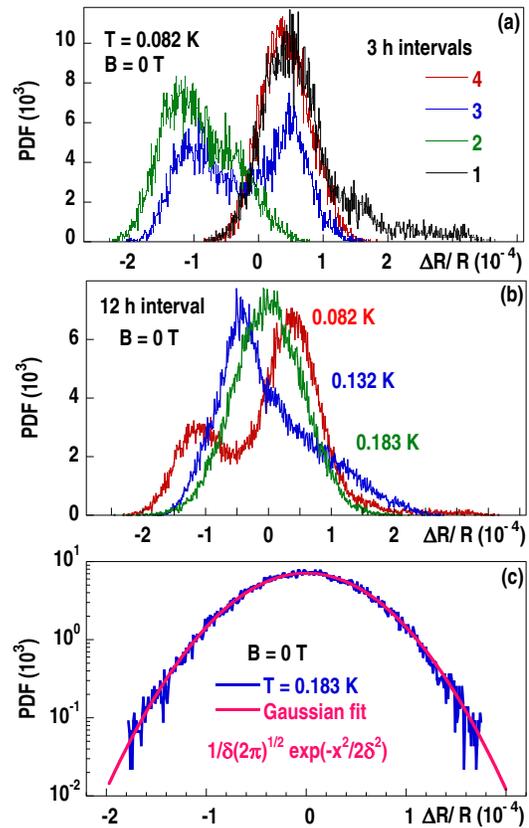}
\caption {(Color online) (a) Probability density function (PDF) \textit{vs.} $\Delta R(t)/\langle R\rangle$ for four sequential 3-hour time intervals at $T =0.082$~K and in $B=0$. (b) PDF \textit{vs.} $\Delta R(t)/\langle R\rangle$ for 12-hour intervals at several $T$ with $B=0$.  (c) PDF values at $T=0.183$~K fitted to a Gaussian distribution, as shown ($\delta=5.6\times 10^{-5}$).
}\label{fig:LSCOabHisFFT}
\end{figure}

In particular, at the lowest $T$ [\textit{e.g.} Fig.~\ref{fig:LSCOabNoiseZeroField}(b)], the noise is not Gaussian.  This is clearly seen already in the histograms of $\Delta R(t)/\langle R(t)\rangle$ values or the probability density function (PDF) of the fluctuations.  In general, the low-$T$ PDF has a complex structure with more than one peak and, moreover, its precise shape depends randomly on the observation time.  This is demonstrated, for example, by the PDFs of the noise data obtained during four sequential 3-hour intervals at the same low $T$ [Fig.~\ref{fig:LSCOabHisFFT}(a)].
%
%
The wandering of the PDF shape with time indicates that different states contribute to the resistance as a function of time, signaling that the system is nonergodic (glassy) on experimental time scales.  Although the PDF broadens when the sampling time is increased to 12 hours, since the system has more time to explore the free energy landscape, the PDF remains non-Gaussian [Fig.~\ref{fig:LSCOabHisFFT}(b)].  When $T$ is increased, on the other hand, non-Gaussian effects become less pronounced [Fig.~\ref{fig:LSCOabHisFFT}(b)], and vanish completely already at $T \sim 0.2$~K where PDF is described by the Gaussian distribution [Figure~\ref{fig:LSCOabHisFFT}(c)].

\begin{figure}
\includegraphics[width=6cm]{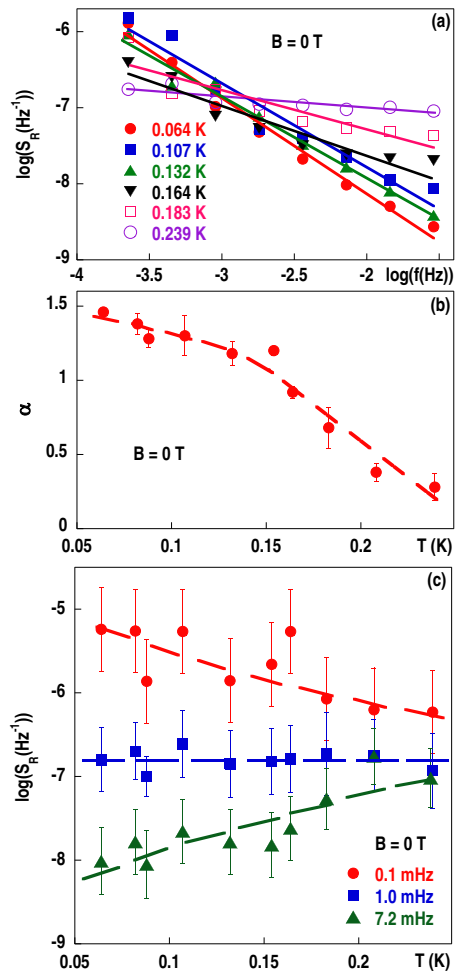}
\caption{(Color online)  (a) Octave-averaged power spectra $S_R(f)$ shown for different $T$ and $B=0$ have been corrected for the white background noise. Solid lines are linear least-squares fits to the form $S_{R} \propto 1/f^{\alpha}$.  (b) The exponent $\alpha$ \textit{vs.} $T$. The dashed line guides the eye.  (c) $S_R(f)$ for $f =0.1, 1.0$, and 7.2~mHz, determined from the fits in (a), as a function of $T$. The error bars are standard deviations of the data.  As $T$ is reduced, (high-) low-frequency contributions to conductivity become (decreasingly) increasingly important.}\label{fig:PowerAlfaBZero}
\end{figure}

The normalized power spectra $S{_R}(f)$ (where $f$ is the frequency) of the noise $\Delta R(t)/\langle R\rangle$
obey the well-known empirical law $S_{R}\propto 1/f^{\alpha}$ [Fig.~\ref{fig:PowerAlfaBZero}(a)].
%
%
The exponent $\alpha$ increases as $T$ is reduced [Fig.~\ref{fig:PowerAlfaBZero}(b)], indicating a shift of the
spectral weight toward lower $f$ and the slowing down of the dynamics.  A careful analysis of the noise magnitude $S_R(f)$ at fixed $f$ also shows that slow dynamic contributions (\textit{i.e.} low $f$) to conductivity become increasingly important with decreasing $T$ [Fig.~\ref{fig:PowerAlfaBZero}(c)].  Moreover, large values of the exponent $\alpha$ at low $T$ ($\alpha>1$, up to $\approx 1.5$) reflect the increasing non-Gaussianity of the noise as $T \rightarrow 0$.  Similar behavior was observed in the $R_c$ noise,\cite{Raicevic08} as well as in some other charge\cite{Bogdanovich02,Jaroszynski02} and spin\cite{Jaroszynski98,Neuttiens00} glasses.  However, we note that, while the in-plane noise becomes nearly white (\textit{i.e.} $\alpha\approx 0$) already at $T \sim 0.2$~K, the out-of-plane noise retains its $1/f$ frequency dependence ($\alpha\approx 1$) even at $T\sim 0.3$~K.\cite{Raicevic08}  This difference, together with the absence (presence) of switching events in the in-plane (out-of-plane) noise, suggests, not too surprisingly, the existence of some additional processes that contribute to the $c$-axis transport and noise.  Indeed, the in-plane value of the power spectral density is approximately two orders of magnitude smaller than the out-of-plane value.\cite{Raicevic08}

\subsection{Noise in magnetic field}

\begin{figure}
\includegraphics[width=8cm]{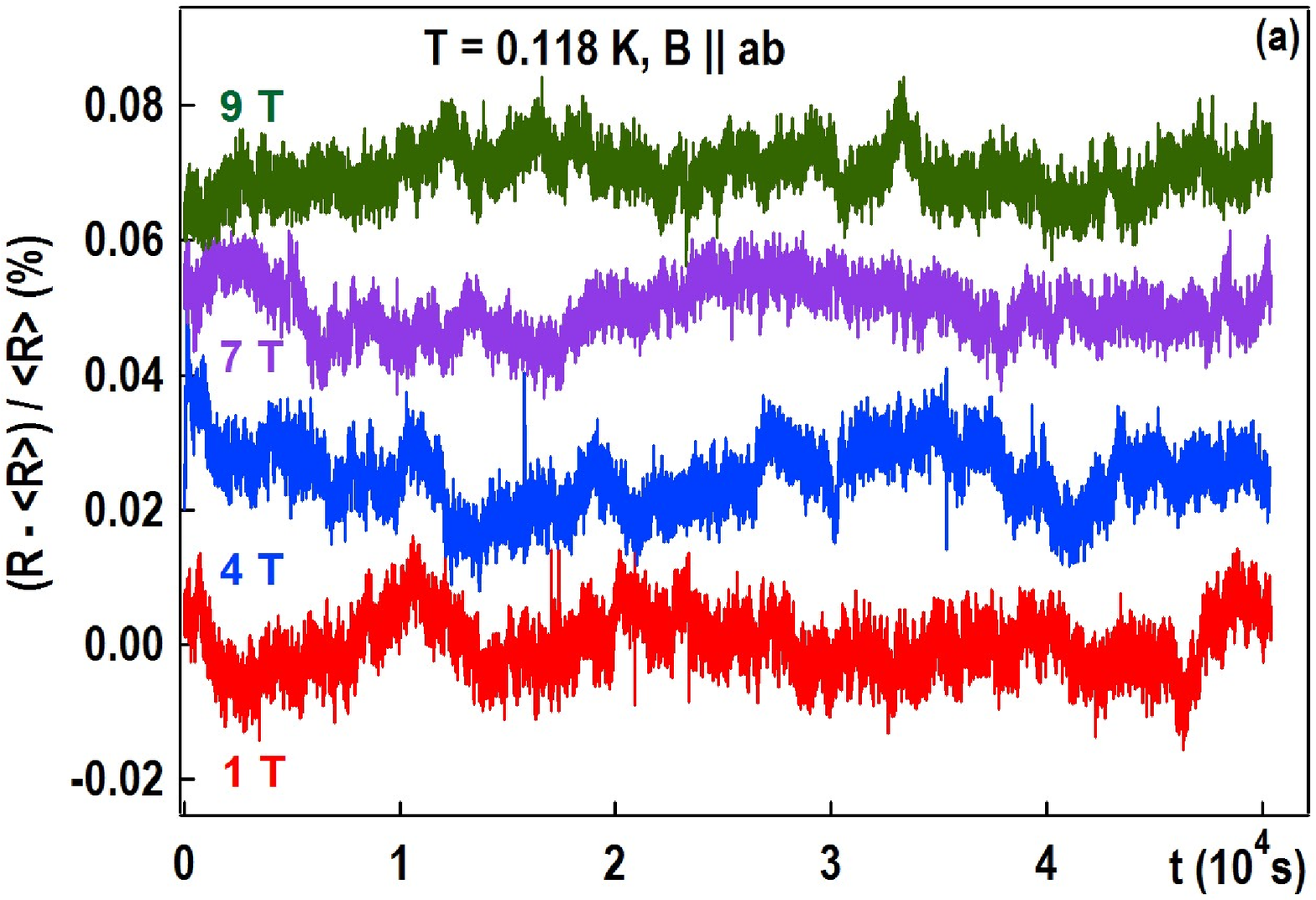}\\
\includegraphics[width=6.5cm]{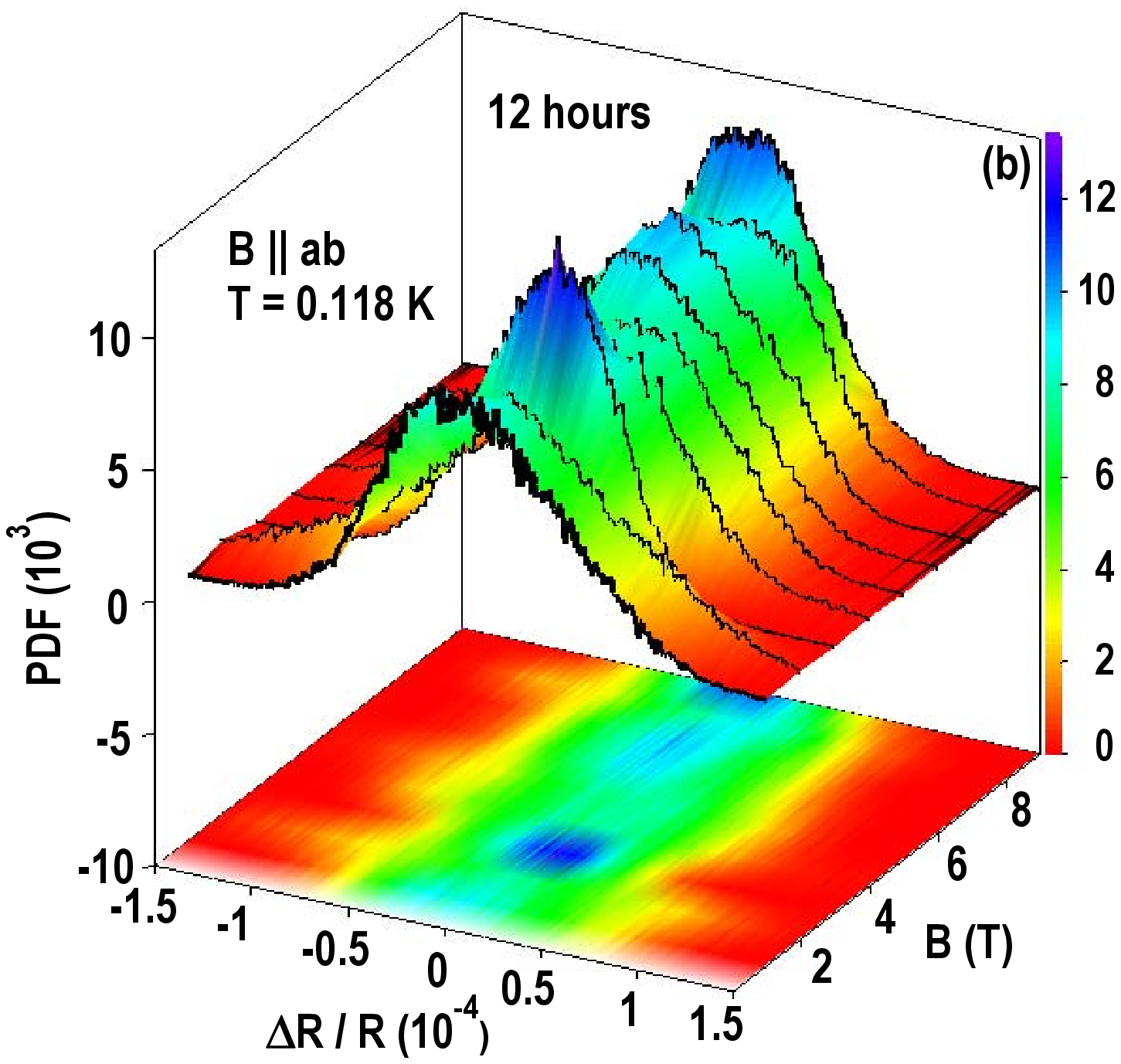}
\caption{(Color online)  (a) Noise $\Delta R(t)/\langle R\rangle$ \textit{vs.} time at $T =0.118$~K for several $B\parallel ab$.  All traces are shifted vertically for clarity.  (b) PDF \textit{vs.} $\Delta R/R$ in different fields for 12-hour observation periods; $T=0.118$~K.}\label{fig:LSCOabNoiseInFieldHisPowerAlfa}
\end{figure}

\begin{figure}
\includegraphics[width=6cm]{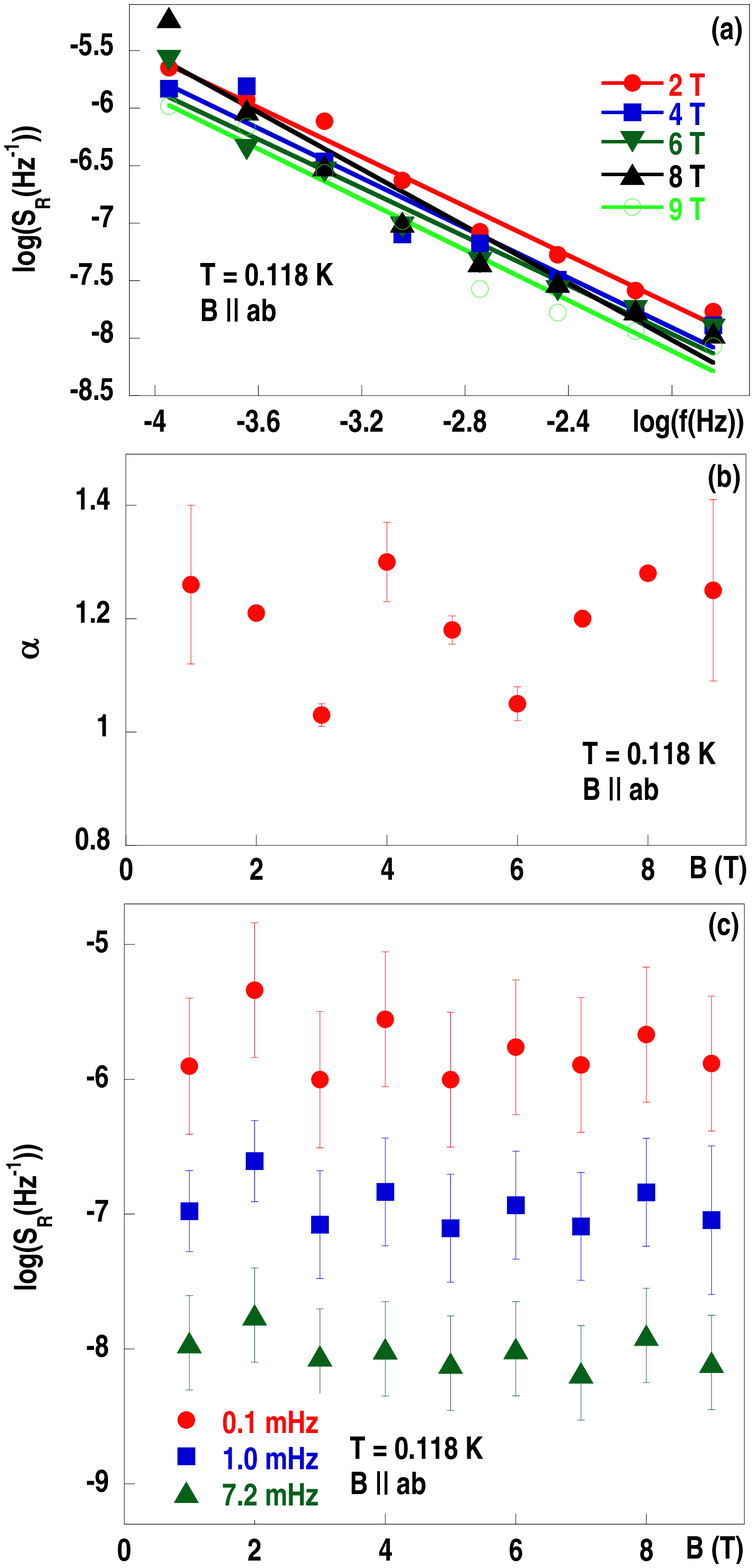}
\caption{(Color online)  (a) The octave-averaged power spectra $S_R(f)$, shown for several $B$ at $T=0.118$~K, have been corrected for the white background noise. Solid lines are linear least-squares fits to the form $S_{R} \propto 1/f^{\alpha}$.  (b) The exponent $\alpha$ \textit{vs.} $B$ at $T=0.118$~K.  (c) $S_R(f)$ for $f = 0.1, 1.0$ and 7.2~mHz, determined from the fits in (a), \textit{vs.} $B$ at $T=0.118$~K. The error bars are standard deviations of the data.}\label{fig:PowerAlfa}
\end{figure}

The relative changes of $R$ with time have been measured also in magnetic fields $B\parallel ab$ of up to 9~T at a fixed, low $T$ [Fig.~\ref{fig:LSCOabNoiseInFieldHisPowerAlfa}(a)], where the zero-field noise is non-Gaussian.
%
%
Similar to the $c$-axis case,\cite{Raicevic08} the data do not seem to show any effect of the magnetic field on either the amplitude or the character of the noise.  For example, the low-$T$ PDF remains non-Gaussian [Fig.~\ref{fig:LSCOabNoiseInFieldHisPowerAlfa}(b)] up to the maximum applied field.  The noise power spectra $S_{R} \propto 1/f^{\alpha}$ do not show any effect of $B$ either [Fig.~\ref{fig:PowerAlfa}(a)]: the large low-$T$ values of the exponent $\alpha$ do not vary with $B$ [Fig.~\ref{fig:PowerAlfa}(b)], and the magnitude
%
%
of the noise is independent of $B$ in the entire experimental range of frequencies [Fig.~\ref{fig:PowerAlfa}(c)].
Furthermore, while the resistance itself depends strongly on the cooling procedure,\cite{Ivana-pMR} all the noise characteristics are independent of the magnetic history.  In particular, the same results are obtained after field-cooling and zero-field cooling protocols, again in analogy with Ref.~\onlinecite{Raicevic08}.

\subsection{Higher-order statistics}

Many noise sources in nature are Gaussian.  For Gaussian random processes in general, all $n$-point ($n>2$) correlation functions can be expressed in terms of products of pairwise correlation functions.\cite{Weissman88,Kogan-book}  Therefore, the measurement of higher-order correlations cannot give any new information beyond that which is already contained in the pairwise correlation function (the Fourier transform of the power spectrum).

A superposition of many independent sources gives rise to a Gaussian noise.  For example, noninteracting, identical sources or fluctuators (\textit{e.g.} electron emission from a cathode) produce noise that is random, not correlated in time, so that the corresponding power spectrum is independent of frequency (``white'').  On the other hand, a superposition of many noninteracting two-state systems (TSSs) with a broad distribution of characteristic times gives rise to a Gaussian noise with a power spectrum of $1/f$ type.  In such a material, the spectrum $S$ at each $f$ is given by the sum of fixed components from different TSS.  Repeated measurements should thus yield the same values of the power spectrum at a given $f$, within experimental accuracy.  In reality, since actual measurements take place over finite times, repeatedly measured $S$ fluctuates randomly around its mean value, which is time-independent.  Thus the power spectrum of the time series of such measurements of $S$ (the ``second spectrum'') is white; that is, the noise is Gaussian.

In many complex systems, however, strong deviations of the noise statistics from Gaussianity have been observed, indicating that the noise cannot be described by noninteracting models.\cite{Weissman88,Kogan-book,Weissman93}  For example, repeated measurements reveal the ``wandering'' of the power spectrum at a given $f$ with time, rather than settling into a sum of fixed components coming from different fluctuators.\cite{Weissman93}  This wandering means that, during the time scale of each measurement, the system ``visits'' only a small number of states in the free energy landscape (\textit{e.g.} several shallow valleys within one deep valley).  In that case, the value of $S$ at a given $f$ is determined by the characteristic relaxation times of the shallow valleys.  After a sufficiently long time, the system undergoes a transition to another deep valley with a different set of relaxation times for transitions between its shallow valleys.  Repeated measurements of $S$ at a given $f$ will thus yield values that are correlated in time, with the corresponding nonwhite second spectrum.  This type of system is obviously non-ergodic.

The use of higher-order correlation and response functions is now broadly accepted as a standard tool for analyzing dynamical heterogeneities in a wide variety of systems.  In particular, the second spectrum, a fourth-order noise statistic, was first used to investigate the dynamics of spin glasses and provide information about correlations among local fluctuators.\cite{Weissman92,Weissman93} The second spectrum $S_{2}(f_{2},f)$ is defined\cite{Weissman88} as the power spectrum of the fluctuations of $S_{R}(f)$ with time and, in practice, is calculated in narrow frequency bands (\textit{e.g.} octaves) $f=(f_{L},f_{H})$.  It was introduced as a technique for understanding how noise power redistributes among narrow frequency bands as a system evolves in time. The second spectrum's utility is based on its sensitivity to correlations among the fluctuators that give rise to the non-Gaussian properties of the noise.  As discussed above, if the noise arises from an ensemble of statistically independent fluctuators (Gaussian noise), $S_{2}(f_{2},f)$ is white (\emph{i.e.} independent of $f_{2}$).  On the other hand, $S_{2}(f_{2}, f)\propto 1/f_{2}^{1-\beta}$ for interacting fluctuators.\cite{Weissman88, Weissman92, Weissman93}

\begin{figure}
\includegraphics[width=7cm]{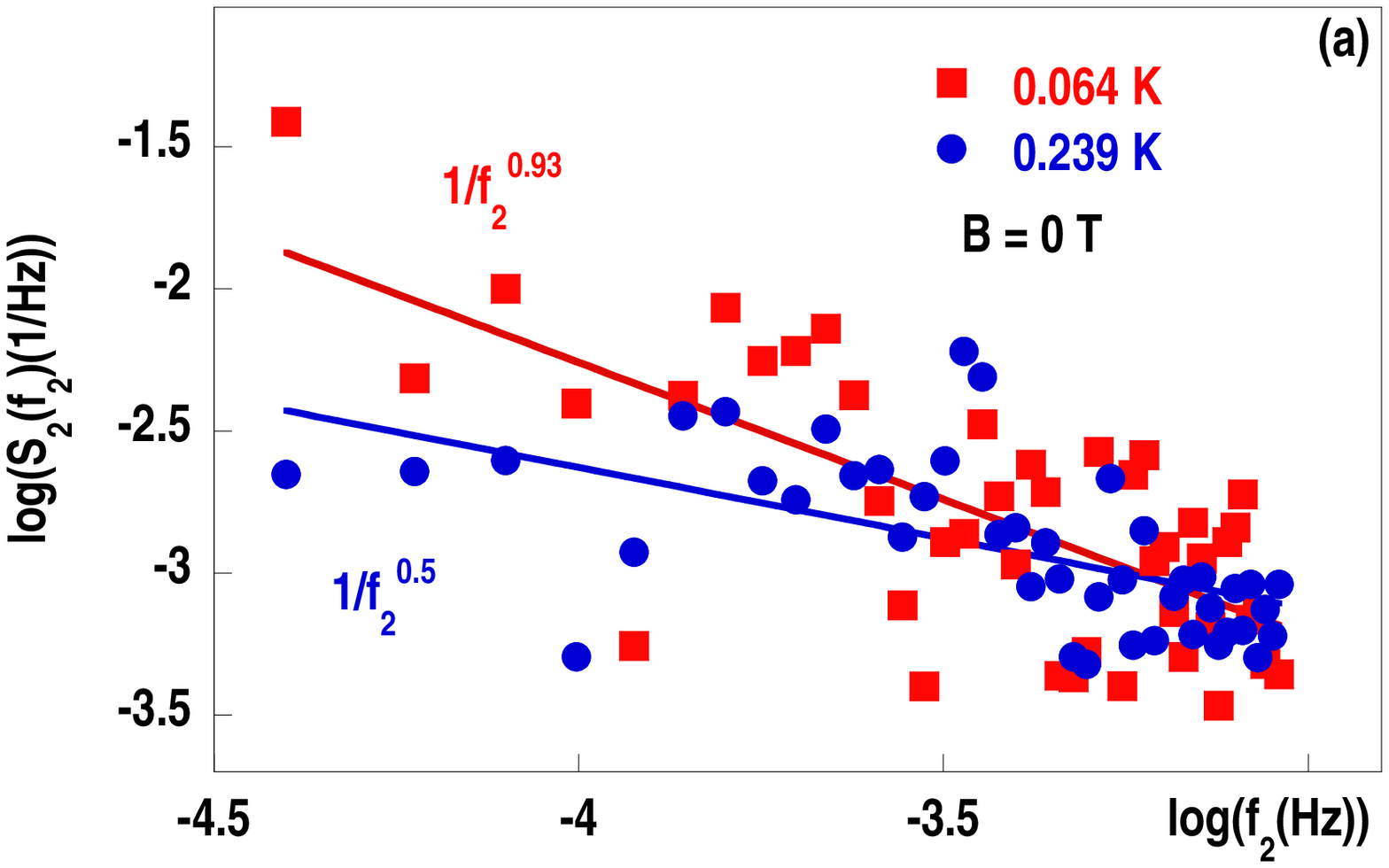}\\
\includegraphics[width=7cm]{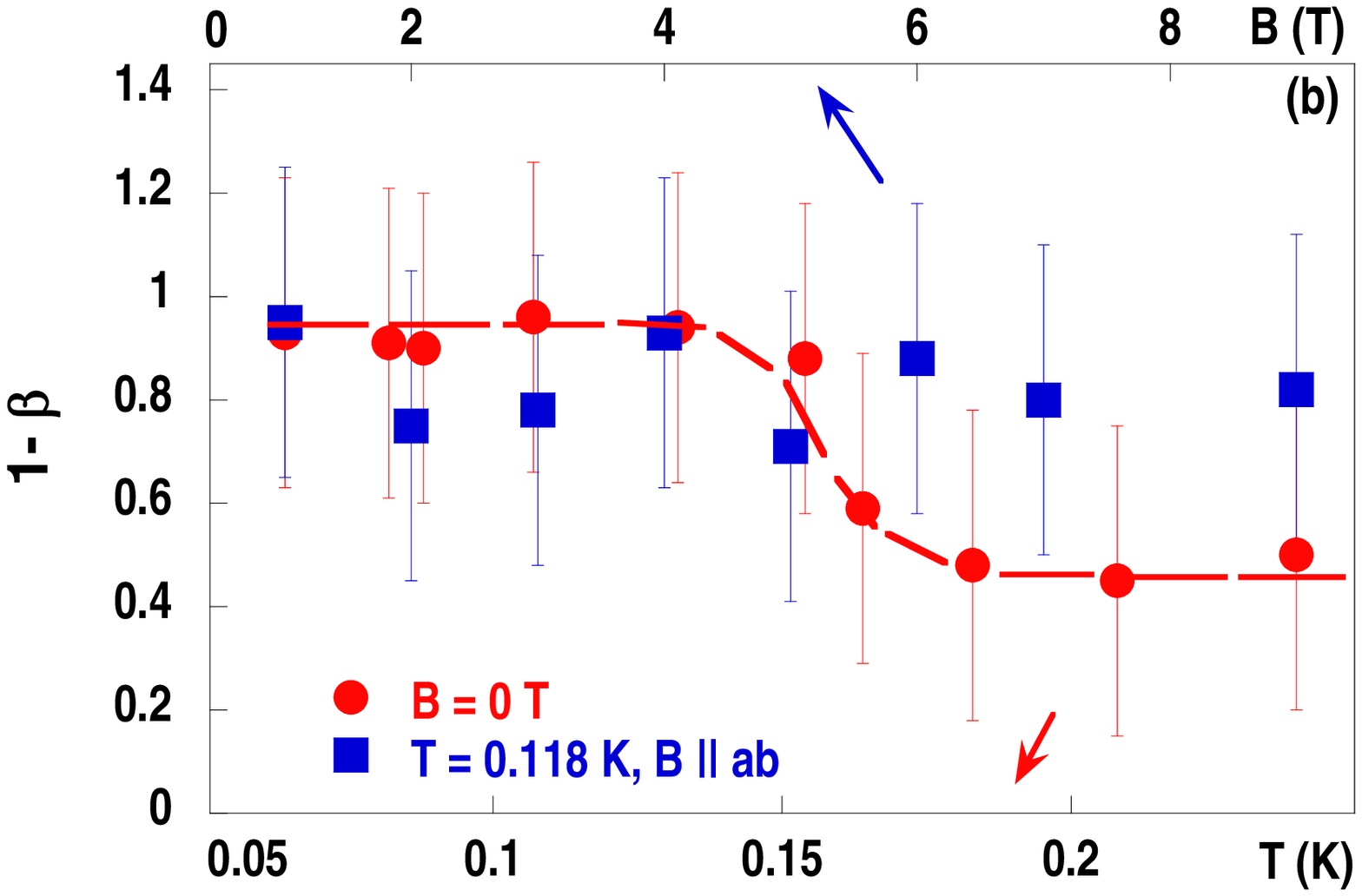}
\caption{(Color online)  (a) Normalized second spectra $S_{2}(f_{2})$, with the Gaussian background
subtracted, for two $T$ in $B = 0$. Solid lines are fits to $S_{2}\propto 1/f_{2}^{1-\beta}$. (b) The exponent ($1-\beta$) as a function of $T$ ({\Large $\bullet$}; bottom scale) and $B$ ($\blacksquare$; top scale). The error bars show the maximum standard deviation of the data. The dashed line guides the eye.  The results shown in (a) and (b) were obtained by averaging over the 0.5-1, 1-2, and 2-4~mHz octaves.}\label{fig:LSCOabSStwoT1MinusBetavsBvsT}
\end{figure}

The second spectra have been calculated using digital filtering.\cite{Seidler96, Abkemeier97} First, the fast Fourier transform (FFT) is applied to the data, which are then digitally bandpass filtered in a given frequency range $f =(f_{L}, f_{H})$ with $f_{H}=2f_{L}$. This allows the analysis of $\Delta R(t)/\langle R\rangle$ in any bandpass. After taking the inverse FFT of the filtered data, they are squared point by point to obtain the variance of the noise \textit{vs.} time for a given $f$. Finally, the power spectrum of the time-dependent variance of the noise signal
($\emph{i.e.}$ the second spectrum) is calculated. $S_{2}(f_2,f)$ is then divided by the square of the averaged total power over unit time in the bandwidth limited $S_{R}(f)$ to obtain the normalized second spectrum.
Figure~\ref{fig:LSCOabSStwoT1MinusBetavsBvsT}(a) illustrates the
%
%
results of that analysis.  The normalized second spectra, with the Gaussian background subtracted, are shown for the lowest and highest $T$ in the experiment.  The exponent $(1-\beta)$, which is a measure of correlations, clearly increases with decreasing temperature. Therefore, as $T$ is reduced, the fluctuators become strongly correlated and the non-Gaussian nature of the noise becomes more pronounced.  A detailed dependence of $(1-\beta)$ on both $T$ and $B$ is given in Fig.~\ref{fig:LSCOabSStwoT1MinusBetavsBvsT}(b).  Just like other noise characteristics, $(1-\beta)$ exhibits $B$-independent behavior at a fixed $T$, indicating that the magnetic field does not affect the nature of the correlations.

The second spectrum has been useful also in distinguishing between different kinetic models that have identical spectra. In particular, the scaling of $S_2$ with respect to $f$ and $f_2$ has been used in studies of spin\cite{Weissman92, Weissman93} and Coulomb glasses\cite{Bogdanovich02,Jaroszynski02,Jaroszynski04} to discriminate between two rival theoretical approaches - the interacting droplet model and a picture of hierarchical dynamics.  In the droplet approach,\cite{Fisher88nonequ,Fisher88equ} which relies crucially on the finite range of the interactions and assumes compact droplets, the slow events (\textit{i.e.} low-$f$ noise) originate from rearrangements (``flipping'') of a smaller number of large droplets, while the higher-$f$ events come from a larger number of smaller elements that are faster.\cite{Weissman92, Weissman93}  Furthermore, according to this picture, large droplets are more likely to interact than small ones with a probability depending on their actual size, leading to the conclusion that $S_{2}$ will be stronger for lower $f$.  However, second spectra obtained for different $f$ need to be compared for a fixed $f_{2}/f$ (\textit{i.e.} on time scales determined by the timescales of the fluctuations being measured), since spectra taken over a fixed time interval average the high-frequency data more than the low-frequency data. Hence, in the interacting droplet model, $S_{2}(f_2,f)$ should be a decreasing function of $f$ at constant
$f_{2}/f$. In the hierarchical picture of glasses,\cite{Binder86} on the other hand, there is no characteristic scale: the second spectra depend only on the ratio $f_{2}/f$, not on $f$, and thus can be collapsed onto a single curve.  Such scale invariant $S_2(f_2,f)$, consistent with the hierarchical picture of glassy dynamics, have been reported in systems with long-range interactions, such as conventional spin glasses (\textit{e.g.} CuMn) \cite{Weissman92, Weissman93} and 2D Coulomb glasses.\cite{Bogdanovich02,Jaroszynski02,Jaroszynski04}

\begin{figure}
\includegraphics[width=7.5cm]{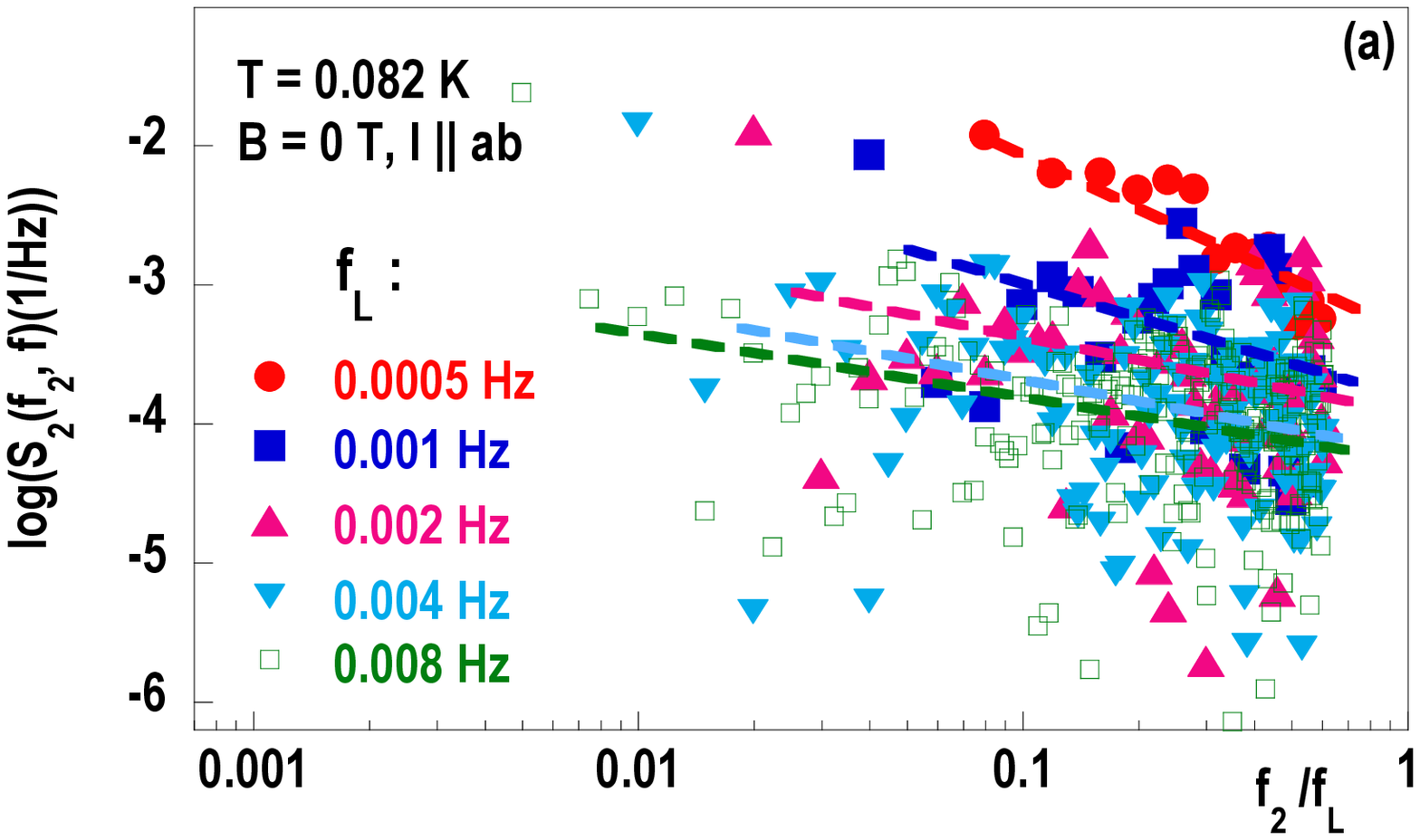}\\
\includegraphics[width=7.0cm]{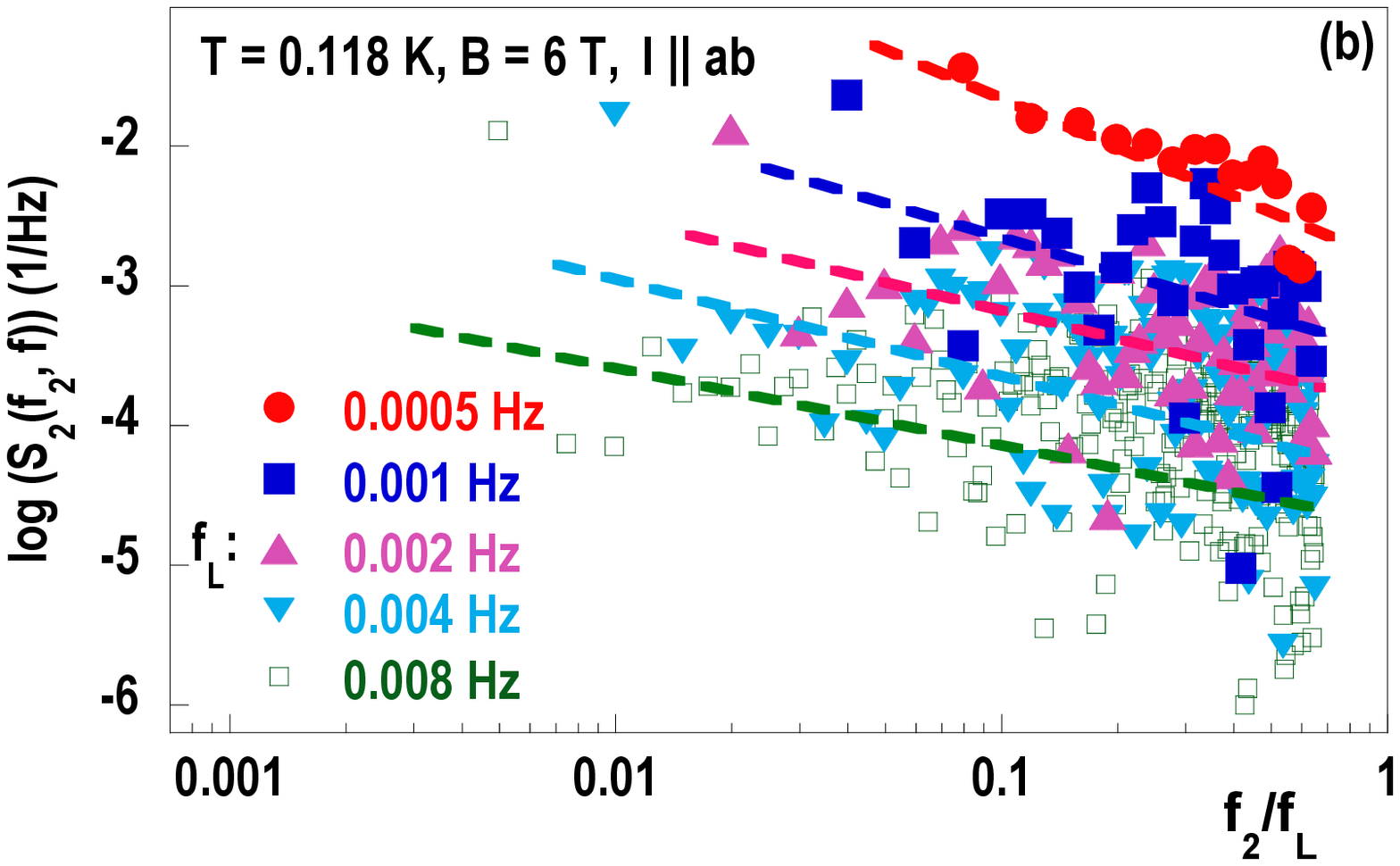}
\caption {(Color online) Second spectrum $S_2(f_2,f)$ measured in octaves $f=(f_{L}, 2f_{L})$ (a) in zero magnetic field at $T=0.082$~K and (b) in $B=6$~T at $T=0.118$~K.  Dashed lines are linear fits to guide the
eye.  No dependence on the magnetic field has been observed within the scatter of data.
}\label{fig:LSCOabScalingZeroField6TField}
\end{figure}

Our findings for the second spectra obtained for different values of $f$ in both zero and finite magnetic fields are shown in Figs.~\ref{fig:LSCOabScalingZeroField6TField}(a) and \ref{fig:LSCOabScalingZeroField6TField}(b),
%
%
respectively.  It is apparent that $S_{2}$ decreases with increasing $f$ for a fixed $f_{2}/f$, consistent with generalized models of interacting, compact droplets or clusters.  Therefore, this result further supports the picture of spatial segregation of holes into interacting, hole-rich droplets or clusters, which are separated by hole-poor AF domains.  These \textit{dynamic} charge inhomogeneities thus result from the competition of short-range interactions, revealed by $S_2(f_2,f)$ (Fig.~\ref{fig:LSCOabScalingZeroField6TField}), and a long-range Coulomb interaction.

\section{CONCLUSIONS}

Our measurements of the in-plane resistance noise in $x=0.03$ LSCO have demonstrated the emergence of slow, correlated dynamics and nonergodic behavior at very low $T\ll T_{SG}$.  The gradual enhancement of this glassy behavior with decreasing temperature strongly suggests that the phase transition to a charge glass state occurs at $T=0$.  Several findings in our study can be used to rule out spins as the origin of the observed glasslike behavior.  For example, the resistance noise spectroscopy reveals that the significant arrest of the charge motion in CuO$_2$ planes occurs only at $T < 0.2$~K (\textit{i.e.} deep inside the spin-glass phase).  Furthermore, while both in-plane and out-of-plane resistance are strongly affected by the magnetic field,\cite{Raicevic08,Ivana-pMR} all the noise characteristics are insensitive to both the magnetic field and the magnetic history, further indicating that the observed glassiness reflects the dynamics of charge, not spins.  This is analogous to the magnetic insensitivity of the noise in a glassy, fully spin-polarized 2D electron system.\cite{Jaroszynski04}  We note that in conventional spin glasses, by contrast, all the resistance noise characteristics are affected by $B$.\cite{Izraeloff89, Jaroszynski98, Neuttiens00}

Therefore, this study provides evidence for the existence of a cluster charge glass ground state in $x=0.03$ LSCO,
in agreement with the conclusions based on noise measurements on $c$-axis oriented samples\cite{Raicevic08}, dielectric\cite{Jelbert08}, and magnetotransport\cite{Ivana-pMR} studies.  The strong non-Gaussianity of the noise indicates that the dynamics of charge clusters cannot be attributed merely to the switching between frozen and non-frozen states of individual clusters (TSS) with different characteristics, but rather that the charge clusters are correlated.  Although they affect the interlayer or $c$-axis transport,\cite{Raicevic08,Ivana-pMR} here we have shown that the correlated charge clusters are located in CuO$_2$ planes, where they seem to coexist with hole-poor AF domains that remain frozen at very low temperatures $T\ll T_{SG}$, at least on experimental time scales.  Further work is needed to examine the evolution of this dynamically inhomogeneous state with doping and its possible coexistence with HTSC.

\acknowledgments
We are grateful to J. Jaroszy\'nski for useful discussions.  This work was supported by NSF DMR-0403491 and DMR-0905843, NHMFL via NSF DMR-0654118, MEXT-CT-2006-039047, EURYI, and the National Research
Foundation, Singapore.

\end{document}